\documentstyle[fleqn,english,epsf]{l-aa}

\def\today{\ifcase\month\or
 January\or February\or March\or April\or May\or June\or
 July\or August\or September\or October\or November\or
 December\fi\space\number\day, \number\year}

\def\todmy{\number\day\space\ifcase\month\or
 January\or February\or March\or April\or May\or June\or
 July\or August\or September\or October\or November\or
 December\fi\space\number\year}

\begin{document}

\selectlanguage{english}

\newcommand{\marc} {mag.arcsec$^{-2}$}
\newcommand{\half} {{{1}\over{2}}}
\newcommand{\ehalf} {{1/2}}
\newcommand{\thalf} {{3/2}}
\newcommand{\chalf} {{5/2}}
\newcommand{\pr} {\partial}
\newcommand{\bdisp} {\begin{displaymath}}
\newcommand{\edisp} {\end{displaymath}}
\newcommand{\beqn} {\begin{equation}}
\newcommand{\eeqn} {\end{equation}}
\newcommand{\beqr} {\begin{array}}
\newcommand{\eeqr} {\end{array}}
\newcommand{\nonu} {\nonumber}
\newcommand{\tal}{\it et al \rm}
\newcommand{\AAA}{\it {Astron. and Astrophys.} \rm}
\newcommand{\AAS}{\it {Astron. and Astrophys. Suppl.} \rm}
\newcommand{\AARev}{\it {Astron. and Astrophys. Rev.} \rm}
\newcommand{\ApJ}{\it {Astrophys. J.} \rm}
\newcommand{\ApJL}{\it {Astrophys. J. Let.} \rm}
\newcommand{\ApJS}{\it {Astrophys. J. Suppl.} \rm}
\newcommand{\AJ}{\it {Astron. J.} \rm}
\newcommand{\AR}{\it {Ann. Rev. Astron. Astrophys.} \rm}
\newcommand{\BAA}{\it {Bulletin of the American Astron. Soc.} \rm}
\newcommand{\MN}{\it {Monthly Notices of the Royal Astron. Society} \rm}
\newcommand{\PASP}{\it {Publ. Astr. Soc. Pac.} \rm}
\newcommand{\PASJ}{\it {Publ. Astr. Soc. Jap.} \rm}
\newcommand{\Nat}{\it {Nature} \rm}
\newcommand{\AnA}{\it {Ann. Astrophys.} \rm}
\newcommand{\ch}{\bf {...check this...} \rm}
\def\RR{{I\!\!R}}
\def\NN{{I\!\!N}}
\def\CC{{I\!\!\!\!C}}
\def\QQ{{l\!\!\!Q}}
\def\R{{\bf R}}
\def\Q{{\bf Q}}
\def\N{{\bf N}}
\def\C{{\bf C}}

\jot=4mm
\mathindent=1mm

\baselineskip=5.0mm

   \thesaurus{**         
	      (**.**.*;  
	       **.**.*;  
	       **.**.*;  
	       **.**.*;  
	       **.**.*;  
	       **.**.*)} 
   \title{On the calculation of the linear stability parameter of periodic
orbits} 
 
 
   \author{C. Barber\`a$^{1,2}$ \& E. Athanassoula$^{2}$} 
 
   \offprints{C. Barber\`a} 
 
   \institute{(1)Dep. Inform\`atica. Escola T\`ecnica Superior  
	      d'Enginyeries. Univertitat Rovira i Virgili,\\ 
	      43006 Tarragona, Spain\\
	      (2) Observatoire de Marseille, 2 place Le Verrier,\\
	      F-13248 Marseille Cedex 4, France
	     } 
 
   \date{Received 5 3, 1998; accepted 26 5, 1998} 
 
   \maketitle 

\begin{abstract}

In this paper we propose an improved method for calculating H\'enon's
stability parameter, which is based on the differential of the
Poincar\'e map using the first variational equation. 
We show that this
method is very accurate and give some examples where it gives correct
results, while the previous method could not cope. 

\end{abstract}

\section{Introduction}
\indent

Over 30 years ago, in a now seminal paper, H\'enon (1965) introduced the
stability parameter $\alpha$, which distinguished whether a given
periodic orbit is stable or not. 
This distinction is essential since the properties of the two types of orbits
differ considerably. Indeed stable periodic orbits trap regular orbits
around them, while unstable ones trigger chaos. The use of H\'enon's
stability parameter is, however, not limited to that. By allowing us
to find precisely the value of the energy for which a given family
changes from stable to unstable, or vice-versa, it allows us to find
the point from which new families bifurcate, since for a periodic orbit
a transition from stability to instability results in a bifurcation of a
new
stable family, while a transition from instability to stability
introduces a new unstable family.

To calculate stability, H\'enon (1965) approached
the differential of the Poincar\'e map by finite differences. This
technique has been so far widely followed in galactic dynamics (e.g. 
Contopoulos \&
Gr\"osb$\o$l
1989 and references therein). Nevertheless it suffers from a
number of disadvantages. We have found it quite adequate in regular
regions, but found it could not cope with difficult chaotic
regions. For this reason we present here an alternative approach, based on
the differential of the Poincar\'e map using the first variational
equation. We will hereafter refer to it as variational
equation method. The variational equation technique is used in other
domains that need very accurate results, like the study of the
three-body problem, the St\"ormer problem or a few others problems  in
celestial mechanics (e.g. Deprit \& Price 1965; Markellos 1974; Markellos
\&
Zagouras 1977), as well as to obtain the Lyapunov exponents (Benettin et
al.,
1976, 1980; Contopoulos et al., 1978; Udry \& Pfenniger 1988). In this paper
the variational equation will allow us to obtain the differential of
the Poincar\'e map exactly. In section \ref{sec:method} we give a
very brief mathematical justification of this method, which is well
known from other applications, and
then we apply it to the particular case of a two-dimensional system which
is stationary in a rotating frame of reference, a case often encountered in 
galactic dynamics problems. Section \ref{sec:axisym} gives an
example demonstrating the accuracy of the method we propose and
section \ref{sec:bar} another example for which this new method gives
accurate results, while finite differences can not cope. We conclude in
section  \ref{sec:conclusions}.

\section{Method}
\label{sec:method}
\indent

\subsection{Notation}
\label{sec:notation}
\indent

Let us consider an autonomous dynamical system, expressed by the  
ordinary differential equation (hereafter ODE).
$\dot{x}=F(x)$ with $x\in\RR^n$ and where $\dot{x}\equiv\frac{dx}{dt}$.
By a solution of this ODE we will mean a map
$\phi : U\subset\RR\times\RR^n\longrightarrow\RR^n$
 such that $\dot{\phi}(t,x)=F(\phi(t,x))$.

Using the precepts of Poincar\'e (1892) we can reduce the study of
a continuous time system (ODE) to the study of an associated discrete
 time system (map) called the Poincar\'e map, $P:V\subset\Sigma
\longrightarrow \Sigma$; $x \longmapsto \bar{x}=\phi(\tau(x),x),$
 where $\Sigma$ is a hyper-surface perpendicular to the flow
$F$, which we will hereafter refer to as section, $V$ is an open set 
in $\Sigma$, and $\tau(x)$ is the time necessary for 
the point $x$ to return for the first time to the section 
with the same sense of transversal of $\Sigma$.
With this technique the problem of calculating the stability of a 
periodic solution of a ODE is reduced to the problem of calculating the
stability
of a fixed point of a map. For this we must check if solutions
starting {\it close} to a fixed point at a given time remain {\it close}
to 
it for all later times (Lyapunov stability).
We thus compute the Taylor series expansion and we study the linear
 term of such an expansion.

Let $x_0$ be a fixed point of $P$ and $x=x_0+\Delta x_0$ a point in its 
neighbourhood. The Taylor expansion of $P(x)$ is

\begin{equation}
P(x)=P(x_0)+DP(x_0)\Delta x_0 + O({\Delta}^2)
\end{equation}

If we denote $P(x)=x_0+\Delta x_1$, we have the linear relation $\Delta x_1
= DP(x_0)\Delta x_0$.
The eigenvalues of $DP$ will determine the stability. In two
dimensions, if $P$ is an area preserving map, the periodic orbit
$\phi(t,x_0)$ is stable if $|\alpha|<1$, where $\alpha$ is the
stability parameter introduced by H\'enon (1965) and defined as
$\alpha = \frac{1}{2}(a_{11}+a_{22})$ where $DP=(a_{ij})$.

\subsection{Calculation of $DP$}
\label{sec:henon}
\indent

H\'enon (1965) approximated the elements of the Jacobi matrix $DP(x)$
using finite differences, i.e.

\begin{equation}
DP(x)_{ij} = \frac{P_i(x_j+\Delta x_j) - P_i(x_j)}{\Delta x_j} + O(\Delta)
\label{der-hen}
\end{equation}

\noindent
As will be shown below, the above approximation gives sufficient
accuracy in regular regions, but not in regions dominated by chaotic
dynamics.

We can calculate $DP$ exactly, as:
\begin{equation}
DP(x)=\dot{\phi}(\tau(x),x)D\tau(x) + D\phi(\tau(x),x)
\end{equation}

\noindent
The matrix $D\phi$ can be obtained as a solution of the first variational
 equation with initial condition $D\phi(0,x)=Id$.
\begin{equation}
\displaystyle\frac{d}{dt}D\phi(t,x)=DF(\phi(t,x))D\phi(t,x)
\end{equation}

\noindent
To compute $D\tau(x)$ we use the fact that we are on the hyper-surface. 
Defining the hyper-surface $\Sigma\equiv g(x)=0$ 
and differentiating we obtain 
after some algebra: 
\begin{equation}
D\tau(x) = -\frac{1}{(\nabla g(x_1),F(x_1))}Dg(x_1)D\phi(\tau(x),x)
\end{equation}
where $x_1=\phi(\tau(x),x)$ and where the symbol $(\quad ,\quad )$ represents the dot
 product. The denominator is different from zero since the 
hyper-surface $\Sigma$ is perpendicular to the flow $F$.

\subsubsection{Particular case}
\label{subsubsec:Particular case}
\indent

We will now apply the above general technique to a two dimensional
system which is stationary in a rotating frame of reference.
Let us consider an autonomous dynamical system, expressed by the following
ODE
\begin{equation}
\left. \begin{array}{l}
\ddot{x} = -\Phi_x + 2\Omega_b\dot{y} + \Omega_b^2 x \\
\ddot{y} = -\Phi_y - 2\Omega_b\dot{x} + \Omega_b^2 y 
\end{array}\right\}
\end{equation}
where $\Phi(x,y)$ is the potential, $\Omega_b$ is the pattern speed of 
the coordinate system in which the dynamical system is stationary, and 
$\Phi_x$ and $\Phi_y$ denote the partial derivatives of the
potential with respect to $x$ and $ y$ respectively.

Writing the above second order ODE as a system of four first order ODE,
and using the momenta $X = \dot{x} - \Omega_b$ and $Y = \dot{y} +
\Omega_b$ we get
\begin{equation}
\left. \begin{array}{l}
\dot{x} = X + \Omega_b y\\
\dot{X} = -\Phi_x + \Omega_b Y\\
\dot{y} = Y - \Omega_b x\\
\dot{Y} = -\Phi_y - \Omega_b X
\end{array}\right\}
\end{equation}

This can also be expressed in vectorial form $\dot{z}=F(z)$ with
$z=(x,X,y,Y)$ and since below we will need the individual components
of $F$, we will denote them as $F_j$ with $j=1$ to 4. Hereafter,
numbers as subindex indicate a component.

Since on the Poincar\'e section $y=0$ and since the energy in a
rotating frame, $E_j(x,X,y,Y)$, is an integral of motion, we can
express  $Y$ as a function of $x$ and $X$, i.e. $Y=Y(x,X)$. This restricts
the Poincar\'e map to the two-dimensional space $(x,X)$. since, however,
the system of first order ODEs has four equations, the
variational equation and the method to calculate the differential of
the Poincar\'e map hold in a four-dimensional space. We
must therefore express the Poincar\'e map in four dimensions in order to
calculate the differential and finally project in the two-dimensional
space $(x,X)$ as shown schematically below

\begin{equation}
\left.
\begin{array}{ccc}
(x,X,0,Y) & \stackrel{\Psi}{\longrightarrow} & (\bar{x},\bar{X},0,\bar{Y})
\\

\makebox[0pt][r]{$i$} \uparrow & & 
\downarrow\makebox[0pt][l]{$\pi$} \\

(x,X) & \stackrel{P}{\longrightarrow} & 
(\bar{x},\bar{X}) 
\end{array}
\right\}\Rightarrow P=\pi\circ\Psi\circ i
\end{equation}

\noindent
Where $i$ includes $(x,X)$ in $\RR^4$ using the section equation and
the integral of motion, $\Psi$ is the Poincar\'e map in $\RR^4$ and
$\pi$ projects the two first coordinates in $\RR^2$.
\noindent
We can therefore write the differential of the Poincar\'e map as
$DP=D\pi\circ D\Psi\circ Di$.

Let us now proceed as in the general case, taking the derivative of $\Psi(z)$
\begin{equation}
D\Psi(z)=\dot{\phi}(\tau(z),z)D\tau(z) + D\phi(\tau(z),z)
\label{DPsi}
\end{equation}

\noindent
Here $\dot{\phi}(\tau(z),z)=F(\phi(\tau(z),z))$, $D\phi(\tau(z),z)$
 is the solution of the first variational equation with initial
 condition $D\phi(0,z)=Id$. To obtain $D\tau(z)$ we use the section
 equation $y=0$, which in $\RR^4$ is equivalent to $\Psi_3(z)=0$. We
 differentiate it and we obtain that the third vector of the matrix
 equation (\ref{DPsi}) should be equal to zero, i.e. $D\Psi_3(z)=\vec{0}$, 
from which it follows that

\begin{equation}
D\tau(z)=-\frac{D\phi_3(\tau(z),z)}{F_3(\phi(\tau(z),z))}
\end{equation}

Now we can express $DP$ in matrix form 
\begin{equation}
\!\!
{DP}\!=\!\left(\begin{array}{cccc}
1 & 0 & 0 & 0 \\
0 & 1 & 0 & 0 
\end{array}\right)
\!\!\!
\left(\begin{array}{cccc}
\partial_x\Psi_1 & \partial_{X}\Psi_1 & 0 & \partial_Y\Psi_1 \\
\partial_x\Psi_2 & \partial_{X}\Psi_2 & 0 & \partial_Y\Psi_2 \\
0 & 0 & 0 & 0 \\
\partial_x\Psi_4 & \partial_{X}\Psi_4 & 0 & \partial_Y\Psi_4 
\end{array}\right)
\!\!\!
\left(\begin{array}{cc}
1 & 0 \\
0 & 1 \\
0 & 0 \\
Y_x & Y_{X}
\end{array}\right)
\end{equation}

\noindent
Where $Y$ is, as discussed above, considered as a function of $x$ and
$X$. We see that in order to obtain $DP$ we need to apply $D\Psi$ only to
$Di$, 
that is, we need to apply $D\phi$ only to $Di$.
Then, instead of solving the variational equation with initial 
condition $D\phi(0,z)=Id$, which gives us $D\phi(t,z)$, we solve it
 applied to the vectors in $Di$, which means solving $\dot{v}=DF(v)$ 
with initial condition 
$v^0=\left(1,0,0,Y_x\right)^T$ and $w^0=\left(0,1,0,Y_{X}\right)^T$$v^0$.
We will now show how it can be computed in practice.
\newline
We integrate simultaneously the orbit and the variational equation for the two vectors in $Di$.
\begin{equation}
\left. \begin{array}{l}
\dot{x} = X + \Omega_b y\\
\dot{X} = -\Phi_x + \Omega_b Y \\
\dot{y} = Y - \Omega_b x \\
\dot{Y} = -\Phi_y - \Omega_b X 
\end{array}\right\}
 \quad \mbox{with} \quad \phi(0,z_0)=
\left(\begin{array}{c}
x_0 \\
X_0 \\
0 \\
Y_0
\end{array}\right) 
\end{equation}

\begin{equation}
\left(\begin{array}{c}
\dot{v_1} \\
\dot{v_2} \\
\dot{v_3} \\
\dot{v_4}
\end{array}\right)=
\left(\begin{array}{cccc}
0 & 1 & \Omega_b & 0 \\
-\Phi_{xx} & 0 & -\Phi_{xy} & \Omega_b \\
-\Omega_b & 0 & 0 & 1 \\
-\Phi_{yx} & -\Omega_b & -\Phi_{yy} & 0 
\end{array}\right)
\left(\begin{array}{c}
v_1 \\
v_2 \\
v_3 \\
v_4
\end{array}\right)
\end{equation}
with $v=v^0$ and with $v=w^0$

Finally with those vectors we write $DP$ as
\begin{equation}
$$DP=
\left(\begin{array}{cc}
\bar{v}_1^0-\displaystyle\frac{F_1}{F_3}\bar{v}_3^0 &
\bar{w}_1^0-\displaystyle\frac{F_1}{F_3}\bar{w}_3^0 \\
 & \\
\bar{v}_2^0-\displaystyle\frac{F_2}{F_3}\bar{v}_3^0 &
\bar{w}_2^0-\displaystyle\frac{F_2}{F_3}\bar{w}_3^0
\end{array}\right)
$$
\label{DP}
\end{equation}
Remember that $\bar{v}_j$ indicates the first return to the section
- in the same sense of traversal - after the point $v_j$. This method 
involves the integration of a system of
12  
rather than 4 equations, because we don't move only the point, but also
 two vectors. This of course takes more computational time, but it gives 
us an accurate value for $DP$, since the error in $DP$ is
of the same order as that of the section points.

\section{Application to an axisymmetric potential}
\label{sec:axisym}

\par
As a first example let us take the axisymmetric logarithmic potential
\begin{equation}
\Phi_L = 0.5 v_0^2 \ln (R_c^2 + R^2)
\end{equation}

\noindent
where $v_o$ and $R_c$ are constants taken, in our example, to be equal
to $1.$ and $0.1$ respectively (Binney \& Tremaine 1987). The orbits of
the
 main, circular family present no difficulty, so both finite
differences and the variational equation method give roughly the same
results. We can, however, compare the accuracy of the two methods by
calculating the determinant of $DP$, which we will hereafter refer to
as $D$. Its elements for finite differences are given by
eq. (\ref{der-hen}), while for the variational equation method by
eq. (\ref{DP}). 

\begin{figure}
\epsfxsize=2.2truein
\centerline{\epsffile{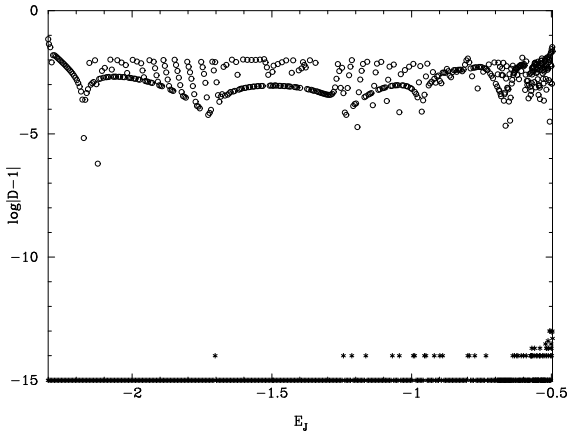}}

\vspace{1.cm}
\caption{
Value of the $\log\vert D-1\vert$ as a function of the Jacobi energy $E_J$
for the axisymmetric logarithmic potential. The values obtained with
finite differences are given by an open circle and those obtained with 
the variational equation method by a star.
}
\label{axi-log}
\end{figure}

A perfectly accurate calculation would of course give $D=1$, and for less
accurate calculations the determinant will deviate more from unity
than for more accurate 
ones. We calculated finite differences for $\Delta x = 10^{-3}$,
$10^{-4}$,
 $10^{-5}$ and $10^{-6}$ and we found that $10^{-3}$ gives the most
accurate
 results. They are compared with those of the variational equation method 
in Fig.~\ref{axi-log}, which shows for both the value of $\log\vert
D-1\vert$.
 The difference in accuracy between the two methods is striking!
The finite difference method gives an accuracy between $10^{-5}$ and $10^{-1}$,
while the accuracy  of the variational equation method is bound only by the
accuracy of the orbit calculation. It is this limiting
accuracy that results also in the ``quantisation'' of the resulting
values. Thus the variational equation method gives, in most cases, an
accuracy of at least $10^{10}$ better than that obtained with finite
differences.
\newline
The improvement in accuracy depends on the case considered and is more 
important in more chaotic regions.

\section{Potential of a barred galaxy}
\label{sec:bar}
\indent

\begin{figure}
\epsfxsize=2.2truein
\centerline{\epsffile{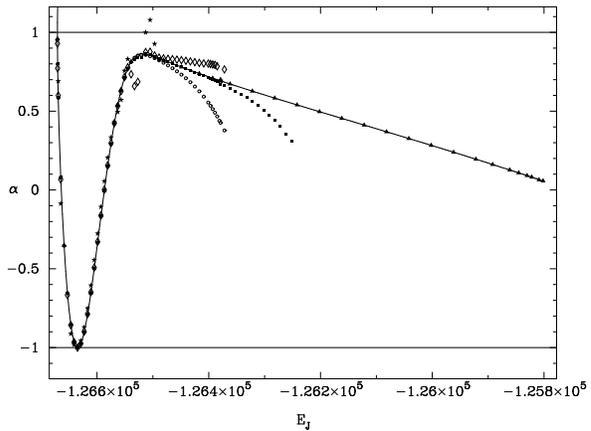}}
\vspace{1.cm}
\caption{
Stability parameter $\alpha$ as a function of the Jacobi energy $E_J$
for a Lagrangian family calculated in the model 1 of A92. The results 
obtained with finite differences and a
$\Delta x$ of $10^{-3}$, $10^{-4}$, $10^{-5}$ and $10^{-6}$ are shown 
respectively by open circles, filled squares, open diamonds and stars.
Results
obtained with the variational equation method are shown with a solid line
and
filled triangles. In order to make the figure clearer we have plotted
symbols only at one out of every five calculated points.
For finite differences we have only plotted points for which 
$|D-1|<.5$. By applying a stricter criterion we can eliminate
successively more points, but in this way we get no information
on the stability of the higher energy orbits.
}
\label{lagr-MN92}
\end{figure}

In the previous section we discussed an example where the variational
equation method gives
quantitatively more accurate results. In other cases, however, the
differences between the two methods can be even qualitative.
\newline
For our second example we will use the model 1 of Athanassoula
(1992, hereafter A92). For most principal families the results of the
two methods are in rough agreement, although the accuracy of the
variational equation method is always better by at least as much as
what we saw in the previous example. However, for families whose
characteristic curve approaches the curve of zero velocity
asymptotically, the differences can be much more important and there
can be disagreement even as to whether a given orbit is stable or
unstable. As an example  let us take a Lagrangian family,
the second one from the right in the lower panel of Fig. 2 
of A92. We calculated its stability using finite
differences as well as using the variational equation method and
compare the results in Fig.~\ref{lagr-MN92}. 

Using the variational equation method we were able to show that this
family is stable for as long as we could follow it. On the other
hand finite differences with $\Delta x=10^{-3}$ find that the
orbits are unstable if $E_J > -126283$, while for $\Delta x=10^{-4}$, 
$\Delta x=10^{-5}$ and $\Delta x=10^{-6}$ the values of the 
Jacobi constant for which the family changes from stable to unstable are
-126150, -126000 and -126500 respectively. Thus finite differences
give results that are qualitatively different from those of the
variational equation method and that depend heavily on the adopted
value of $\Delta x$. With the help of Poincar\'e sections we confirmed
that indeed this family is stable and therefore that it is the
variational equation method that gives the correct result. We next
repeated the calculation again using finite differences but this time
keeping only the points for which the value of the determinant $D$ is
sufficiently close to unity and noted that in this way the erroneous
values disappeared. Although we are thus able to remove erroneous
values, we are not able to find the correct value for the stability
parameter, unless we use the variational equation method.

\section{Conclusions}
\label{sec:conclusions}
\indent
In essence the difference between the two techniques is that finite
differences involve the calculation of $\lim_{\Delta x \to 0}$ for
every element in the Jacobian matrix of the Poincar\'e
map. Numerically it is not possible to calculate the $\lim_{\Delta x
\to 0}$, so we have to calculate this expression for different $\Delta
x$ in order to find the optimum one that gives the best approach to
the limit. It is, however, well known that calculating the derivative
from finite differences can be unstable, and that it may not be
possible to obtain a good approximation of $\lim_{\Delta x \to
0}$. Thus this method may need several trials (i.e. more computing
time), an {\it a posteriori} check of the results, and, in certain
cases, it may not be possible to find an appropriate value of $\Delta
x$.

The new method here calculates the Jacobian matrix accurately and
doesn't depend on any $\Delta x$. It may involve some
extra computing time because it involves the integration of more
differential equations, but this is, more often than not, compensated 
by the fact that it needs only one trial.
Its assets are that it involves no tunable
parameter, needs no {\it a posteriori} checking of the results and its
accuracy is only limited by the method used to find the periodic
orbits. In regular regions the two methods give similar results,
although the variational equation method is always more accurate. In
chaotic or ``difficult'' regions, however, e.g. regions involving 
small stable islands in a chaotic sea, using the variational equation 
method may
prove essential for obtaining the correct value of the stability
parameter. Furthermore, a study of such regions may be very time 
consuming and the 
fact that with the variational equation method one can find precisely 
the bifurcation points of new families can be a big help. We thus 
recommend it for all calculations such as the ones
presented here. 

\parindent=0pt
\def\rr{\par\noindent\parshape=2 0cm 14cm 1cm 13cm}
\vskip 0.7cm plus .5cm minus .5cm

\begin{acknowledgements}
We would like to thank M. H\'enon, P. Patsis, 
D. Puigjaner and J.M. Lopez-Besora for useful discussions, A. Bosma and 
C. Garcia-Gomez for their encouragement, the referee, Daniel Pfenniger, 
for comments that helped improve the presentation of the paper, 
and Jean-Charles Lambert for his cheerful help with setting up our software.
 We would also like to thank the
INSU/CNRS and the University of Aix-Marseille I for funds to develop
the computing facilities used for the calculations in this paper.
E.A. acknowledges support from the Newton Institute during the last revisions of this paper.
\end{acknowledgements}


\end{document}